\documentclass[aip,rsi,reprint]{revtex4-1}

\usepackage{amssymb,amsmath}
\usepackage{graphicx} % support the \includegraphics command and options
\usepackage[colorlinks=true,pdfstartview=Fit]{hyperref}		% add links
\usepackage{color}			% for various colored (gray) things
\usepackage{textcomp}
\begin{document}

\title{Split PID control:  two sensors can be better than one} 

\author{Leith Znaimer}
\noaffiliation

\author{John Bechhoefer}
\email[email: ]{johnb@sfu.ca}
\noaffiliation

%\thanks{}
\affiliation{Dept. of Physics, Simon Fraser University, Burnaby, B.C., V5A 1S6, Canada}

\date{\today}

\begin{abstract}
The traditional proportional-integral-derivative (PID) algorithm for regulation suffers from a tradeoff:  placing the sensor near the sample being regulated ensures that its steady-state temperature matches the desired setpoint.  However, the propagation delay (lag) between heater and sample can limit the control bandwidth.  Moving the sensor closer to the heater reduces the lag and increases the bandwidth  but introduces offsets and drifts into the temperature of the sample.  Here, we explore the consequences of using two probes---one near the heater, one near the sample---and assigning the integral term to the sample probe and the other terms to the heater probe.  The \textit{split-PID} algorithm can outperform PID control loops based on one sensor.
\end{abstract}


\maketitle %\maketitle must follow title, authors, abstract and \pacs

Most experiments and many industrial products need to control the behavior of some parameters.  One common example, both in academia and industry, is temperature control.  While passive control (``insulation") is a good first step, active control can improve performance, particularly at lower frequencies.  To date, the most commonly used control algorithm is also one of the first to have been developed.  Known as proportional-integral-derivative (PID) control \cite{bechhoefer05,astrom08b}, it provides a simple, robust solution to many control problems.

Despite its effectiveness, there is one common tradeoff that usually has to be made in the design of the controller.  For ease of language, we will use the case of temperature control, but the problem is generic.  The tradeoff involves the placement of the measuring device (``probe," or thermometer) relative to the ``actuator"  (heater).  If the probe is place near the sample (the device whose temperature is to be controlled), then the integral term of the PID algorithm will ensure that the sample stabilizes at the desired temperature.  However, the time it takes the heater to deliver heat to the sample will imply a tendency to over-correct for a perturbation.  By the time the heater response has arrived, the perturbation has changed.  The solution is then to lower the controller response (``gain"), which makes the controller slower (``lowers the feedback bandwidth").  To improve the response bandwidth, one can move the probe closer to the heater, but then the temperature at the sample may be different from that desired.  In this Note, we present a modification of the standard PID algorithm that can improve upon this tradeoff.
 
\begin{figure}[ht!]
	 \includegraphics[width=8.0cm]{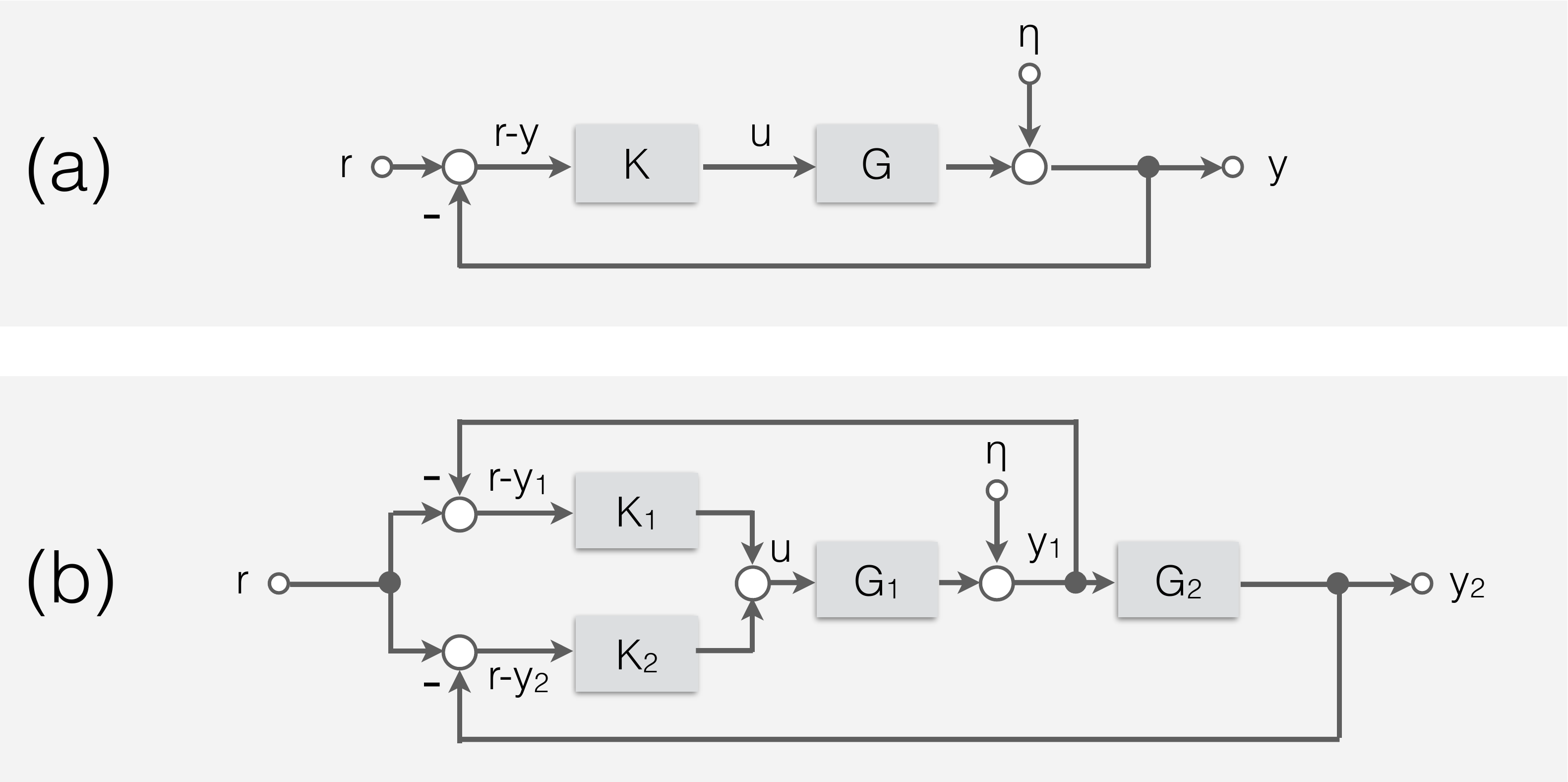}
\caption{Block diagram of control architectures.  (a) Conventional loop to regulate system $G$ against disturbances $\eta$.  The output $y$ is compared to the setpoint reference $r$ and fed back to the cotntroller $K$. (b) Split architecture where two sensor signals $y_1$ and $y_2$ are fed back to two controllers $K_1$ and $K_2$.}
 \label{fig:block} 
 \end{figure}

To understand the motivation for split control more deeply, we first review standard PI control of a single sensor [Fig.~\ref{fig:block}(a)].  We consider a regulator that tries to stabilize the output $y(s)$ at the setpoint, $r$, against disturbances $\eta$.  Here, $y(s)$ is the Laplace transform of a time-domain signal.  Evaluating at $s=i\omega$ gives the frequency response (Fourier transform).   From the block diagram, the input $u = K(r-y)$.  The output $y = Gu + \eta$.  Then, $y= Gu+\eta = GK(r-y) + \eta$, which implies
\begin{align}
%	y= \underbrace{\left( \frac{GK}{1+GK} \right)}_{\equiv T} \, r 
%		+ \underbrace{\left( \frac{1}{1+GK} \right)}_{\equiv S} \, \eta \,,
	y= \left( \frac{GK}{1+GK} \right) \, r + \left( \frac{1}{1+GK} \right) \, \eta  \equiv Tr+S\eta \,,
\end{align}
where $T \equiv y/r$ is the \textit{complementary sensitivity function} and measures the accuracy of the temperature control.  Similarly, $S \equiv y/\eta = 1-T$ is the \textit{sensitivity function} and indicates how much the disturbance $\eta$ affects the output $y$.  Perfect tracking implies $T=1$, and perfect disturbance rejection implies $S=0$.

As an example, we let $G(s) = \tfrac{e^{-\tau(s+\mu)}}{s+\mu}$ be the approximate transfer function between the heater and thermometer probe.  Here, $\tau$ is the lag time, the time it takes heat to propagate the distance between the heater and the temperature probe.  The coefficient $\mu$ represents the distributed loss per distance of heat to the environment.  In fact, the above law is better suited to describing ballistic transport.  For thermal diffusion, a better model would be $G(s) = \tfrac{e^{-\tau(\sqrt{s+\mu})}}{\sqrt{s+\mu}}$, but this choice makes analytic expressions more complicated, without substantially changing the main conclusions.  For controller, we use the proportional-integral (PI) form $K = K_p + K_i/s$.  

Our goal is then to evaluate the frequency response of the functions, $S(i\omega)$ and $T(i\omega)$.  We are interested, in particular, in two aspects.  The first is their DC step response.  For the ability of the controller to reach the desired setpoint, we appeal to the Final-Value Theorem of Laplace transforms\cite{skogestad05} and evaluate $T(s \to 0)$.  In this limit, $G(s) = \tfrac{e^{-\mu}}{\mu} \equiv G_0$, and $K \approx \tfrac{K_i}{s}$.  Then
\begin{align}
	T(s \to 0) = \lim_{s \to 0} \left( \frac{G_0 \frac{K_i}{s}}{1+G_0 \frac{K_i}{s}} \right) = 1 \,.
\end{align}
Because $T \to 1$, the system goes to the desired reference (setpoint) temperature, $r$.  Similarly, $S \to 0$, and there is perfect rejection of slow disturbances, as well.  These desirable outcomes result entirely from integral control. 

We are also interested in the bandwidth of the controller, which, loosely, is the frequency range over which $T \approx 1$ and $S \approx 0$.  As a shortcut to estimating the bandwidth, we find instead the critical gain $K_p^*$ at which the system goes unstable and starts to oscillate.   Since instability results when the denominator of a transfer function vanishes and since both $T$ and $S$ have the same denominator, the bandwidths will be the same in both cases.  (This may not be strictly true if we define bandwidths in terms of specific values of $T$ and $S$, but the scales will be the same.)  Note that instabilities limit the maximum $K_p$ to some fraction of $K_p^*$.  If $K_p$ is too close to $K_p^*$, the response will be weakly underdamped and ``ring down" in an undesirable fashion.

Let us now assume that $\omega^*$ is high enough so that $G \approx  \tfrac{e^{-\tau s}}{s}$ and $K \approx K_p$.  Then $T$ diverges when its denominator $1+GK \approx K_p \tfrac{e^{-\tau s}}{s}/(1+K_p \tfrac{e^{-\tau s}}{s})$ vanishes.  This happens when $K_p \frac{e^{-\tau s}}{s} = -1$.  Since $s=i\omega$, the equation is complex and can be separated into magnitude and phase components.  The magnitude component implies that $\frac{K_p}{\omega} = 1$, or $K_p^* = \omega^*$.  The phase lag should be $\pi$ radians.  The $1/\sqrt{i}$ term  contributes a lag of $\pi/4$ radians, and the remaining phase component, $e^{-i\omega \tau}$ will give an additional lag of $3\pi/4$ radians at instability.  Thus, $\omega^* = \frac{3\pi}{4} \, \frac{1}{\tau}$, implying that the bandwidth (and maximum gain) are both proportional to $\tfrac{1}{\tau}$.  Thus,  placing a sensor close to the sample will lead to a large lag time $\tau$ and low feedback bandwidth.

It is tempting to put the sensor nearer the heater, which will reduce $\tau$ and increase $\omega^*$.  But this strategy leads to temperature offsets and drifts at the sample, whose stability (and accuracy) are, of course, the goal of adding control in the first place.  

Next, consider the situation with two probes [Fig.~\ref{fig:block}(b)].  We represent the system in series $u \to G_1 \to G_2$ to represent having a heater, then probe 1 a short distance away, and then probe 2 and sample a greater distance away.  We thus have $y_2 = G_2 y_1$, $y_1 = G_1 u + \eta$, and $u = -K_1 y_1 - K_2 y_2$.  There are two controllers, $K_1(s)$ and $K_2(s)$.  For the moment, we do not specify their structure.  Notice that the disturbance affects only the intermediate position $y_1$, which is a simplification.  In fact, an outside temperature disturbance is filtered by whatever thermal insulation exists and should affect both $y_1$ and $y_2$, albeit with different frequency responses.  We will assume that the sample is arranged so that the probe at $y_1$ records the disturbance \textit{before} it reaches the sample at $y_2$.  Then corrections applied at $y_1$ can reach the sample at $y_2$  along with the disturbance.  
%  later:  the scheme can also be thought of as a way to introduce feedforward compensation whereby disturbances where we use ``early warning" to compensate for disturbances before they reach the sample.

%In matrix form, the equations for $y_1$, $y_2$, and $u$ are
%\begin{align}
%	\begin{pmatrix} -G_2 & 1 & 0 \\ 1 & 0 & -G_1 \\ K_1 & K_2 & 1 \end{pmatrix} 
%	\begin{pmatrix} y_1 \\ y_2 \\ u  \end{pmatrix}
%	=\begin{pmatrix} 0 \\ 0 \\ K_1+K_2 \end{pmatrix} r
%	+ \begin{pmatrix} 0 \\ 1 \\ 0 \end{pmatrix} \eta \,.
%\end{align}

Defining sensitivity functions $T_1 = y_1/r$ and $T_2 = y_2/r$ and  solving the block-diagram  equations leads to
\begin{align}
	\begin{pmatrix} T_1 \\ T_2   \end{pmatrix}  
		= \frac{G_1(K_1+K_2)}{1+G_1 K_1 + G_1 G_2 K_2}
	\begin{pmatrix} 1 \\ G_2  \end{pmatrix} \,.
\end{align}
%The solutions for $S$ are similar.
%\begin{align}
%	\begin{pmatrix} S_1 \\ S_2 \\ S_u  \end{pmatrix}  = \frac{1}{1+G_1 K_1 + G_1 G_2 K_2}
%	\begin{pmatrix} 1 \\ G_2 \\ K_1 + G_2 K_2 \end{pmatrix} \,.
%\end{align}

To proceed further, we assume that $G_{1,2} =  \tfrac{e^{-\tau_{1,2}(s+\mu_{1,2})}}{s+\mu_{1,2}}$, with the individual values of $\tau$ and $\mu$ depending on the placement of the probes  relative to the heater.  We also specify the controller structure to be the split-PID algorithm, with the PD term acting on $y_1$ and the I term acting on $y_2$.  For simplicity, we neglect the derivative term and set $K_1 = K_p$ and $K_2 = \tfrac{K_i}{s}$.

For the DC behavior, the transfer functions are approximately represented by their DC values $G_1(0)$ and $G_2(0)$.  Then, $s = i\omega \to 0$ implies that 
\begin{align}
	T_2 \approx \frac{G_1(0) G_2(0) (K_p+\frac{K_i}{s})}
		{1+G_1(0) K_p + G_1(0) \, G_2(0) \, \frac{K_i}{s}}
		\to 1 \,,
\end{align}
meaning that the temperature near the heater will match the desired setpoint.  A similar calculation shows $T_1 \to \frac{1}{G_2(0)}$:  
%\begin{align}
%	T_1 \approx \frac{G_1(0) (K_p+\frac{K_i}{s})}
%		{1+G_1(0) K_p + G_1(0) \, G_2(0) \, \frac{K_i}{s}}
%		\to \frac{1}{G_2(0)} \,,
%\end{align}
a setpoint $r$ implies a steady-state heater temperature of $r/G_2(0) > r$:  The heat injected into the sample creates a thermal gradient, because of the thermal resistance (finite conductance) of the material between the heater and the sample.  Putting the integral control on the sample probe nonetheless compensates for this effect. 

We next look at the bandwidth.  Again, all sensitivity functions are governed by the same denominator, $1+G_1 K_1 + G_1 G_2 K_2$.  At high frequencies, $K_2 \approx K_i/s$ becomes small, and the denominator is approximately $1+G_1 K_p$, which is just the result from the single-sensor case.  We conclude that the control bandwidth will be approximately the same as that of a controller that uses a sensor near the heater.

%One issue is that the numerators of the sensitivity functions may also lead to undesired response.  Such problems can arise, for example, when disturbances affect both probes equally.  Then the controller's response to a perturbation at the heater probe may not have the correct phase when it propagates to the sample probe.  Designing the apparatus with the sample better insulated than the heater will help ensure that a response to a perturbation at the heater will continue to be appropriate for the sample when its effect arrives there.  With this qualification, we conclude that split-PID control can improve upon single-sensor control by assuring both accurate temperature at the sample and fast response.  

At this point, we have argued intuitively that split-PID control should improve the performance of a regulator, and we have analyzed a simplified model to support this assertion.  But does it work in practice?  To test the algorithm, we modified a device previously used to measure the capacitance as a function of temperature of a barium-titanate compound that has ferroelectric properties.\cite{bechhoefer07}
 
%Near the Curie temperature of a ferroelectric, there is a large capacitance peak, which is of fundamental interest as an example of a second-order phase transition and of practical interest as the basis for the ceramic capacitors that are ubiquitous in electronic circuits.  (A trillion devices are made each year!\cite{ho10})

The original apparatus used a heater and a nearby temperature probe (thermistor), with  the ferroelectric material a greater distance away from the heater.  Although the temperature control of the thermistor was very good, it was not clear how well it reflected the temperature of the ferroelectric sample itself.  Here, we have built a new device that substitutes a second thermistor in place of the ferroelectric material.  To model the original experiment, we place the ``sample probe" farther from the heater than the ``heater probe."  We then use a data acquisition card (National Instruments, USB-6212) to read the two temperatures and output the calculated power response.  The temperature readings are made by placing each thermistor in a voltage-divider circuit, digitizing at 200 kHz, and averaging 10,000 readings per measurement (RMS noise $< 10^{-3}~^\circ$C).  The algorithm linearizes outputs and inputs to work directly in power and temperature and implements discretized versions of both the standard PID algorithm ($K_p + K_i/s + K_d s$) and our new, split-PID algorithm.  The discretization of the derivative is done via $s \to (1-z^{-1})/T_s$, with sampling time $T_s = 0.2$ s.  A similar approximation is used for the integral.  The operator $z^{-1}$ means ``take the previous signal reading."

\begin{figure}[ht!]
	 \includegraphics[width=8.0cm]{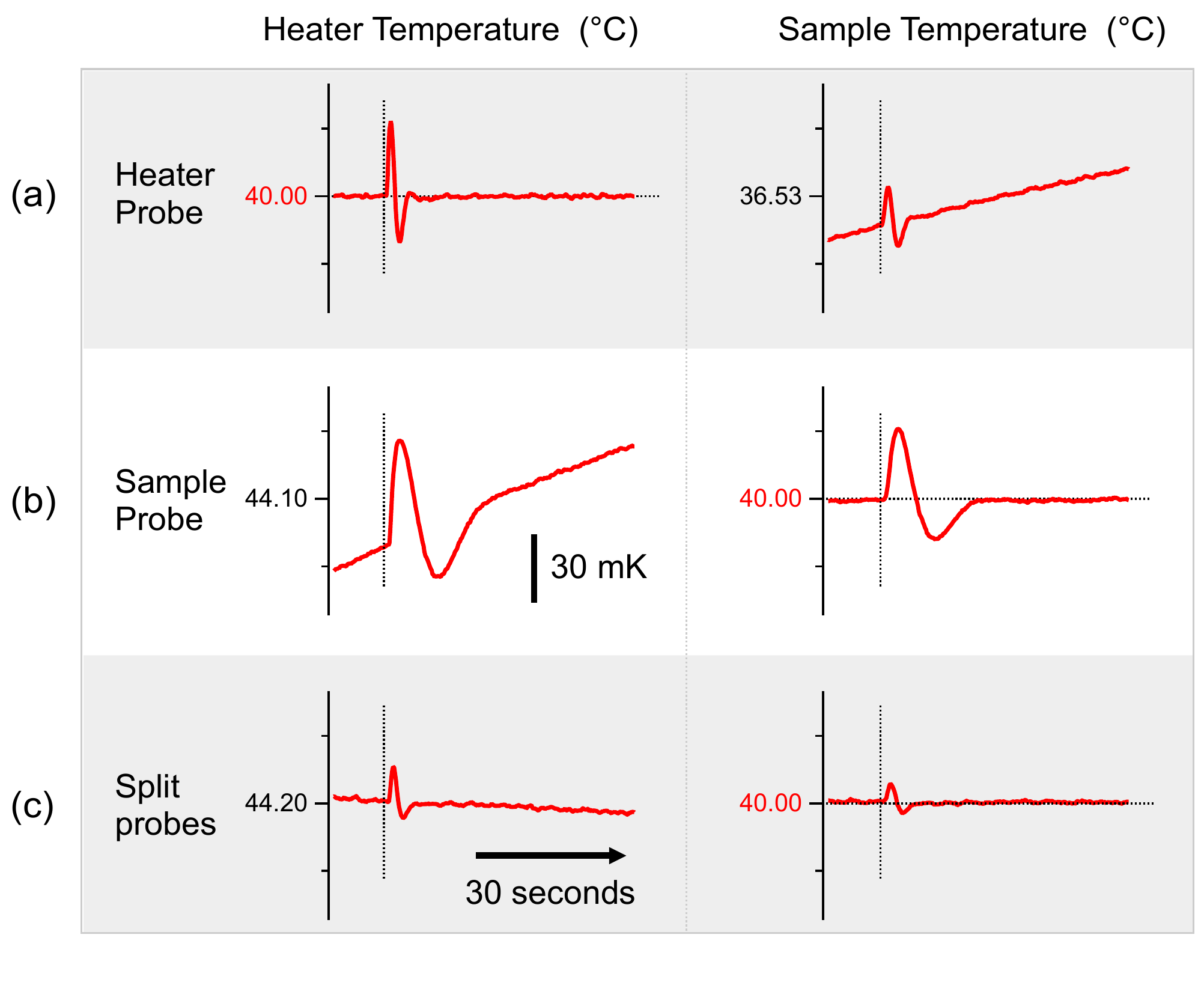}
\caption{Response to a heater pulse perturbation (0.3 W, for 19 ms), as measured by the probe near the heater (left graphs) and by the probe near the sample (right graphs).  The vertical dotted lines indicate the pulse timing.  The control algorithm uses information from (a) the near (heater) probe only  (gains are $K_p=2.3$, $K_i=0.5$, and $K_d=0.6$); (b) the far (sample) probe only (gains are $K_p=0.4$, $K_i=0.05$, and $K_d=0.45$); and (c) both probes (gains are $K_p=2.5$, $K_i=0.5$, and $K_d=0.6$).  The setpoint in all cases was $T=40\,^\circ$C.}
 \label{fig:summary} 
 \end{figure}

Figure~\ref{fig:summary} summarizes the three different types of control.  In Part (a), the control is on the heater probe, with a setpoint of $40\,^\circ$C.  The plot in the left column shows that the heater probe rejects the perturbation; the temperature returns to the setpoint after a few seconds.  By contrast, the sample probe shows a large offset ($3.5\,^\circ$C) and a pronounced drift.  The drift arises because the heat current through the sample varies as the ambient temperature changes.  The temperature is fixed near the heater but must vary near the sample.

In Part (b), PID control is based on the sample probe.  The offset at the sample probe is removed, but the response is much slower.  (Attempting to speed up the response would lead to an underdamped ringing whose envelope would also decay slowly.)  There is offset and drift at the heater, but we do not care about that behavior.

In Part (c), the split-PID control sends ``PD" to the heater probe and ``I" to the sample probe.  The sample temperature is now correct, and the response time is nearly that of the heater probe in Part (a).  

What are the limitations of split-PID control?  First, PID control itself sometimes fails.  For example, PID control does not work well for lightly damped mechanical systems with multiple resonances (e.g., piezoelectric scanners).   But the basic idea of using integral control for the heater sensor and some more general control for the heater sensor may still work.  Indeed, we already do that here when we linearize the sensor and power signals.  
%In addition, one can add rules to deal with cases where the desired setpoint changes are large enough that the controller response saturates (heater is at its upper or lower limits).\cite{astrom08b}

More fundamentally, the two controllers may ``fight" each other, with the actions of one interfering with the other.  In split-PID control, the actions of $K_1$, the PD controller, can take so long to reach the sample that they are out of phase with the perturbation.  The perturbation is properly corrected at the heater---we do not care about that---but not at the sample, where we do care.  The solution is to make disturbances affect the heater probe before they affect the sample.  For example, insulating the sample more than the heater means that disturbances first perturb the heater and then the sample.  Good performance requires paying as much attention to the overall system design as to the controller algorithm.

To put split-PID control in perspective, we note that professional control engineers have long been aware of the potential benefits of adding more sensors (and actuators).  Indeed, MIMO (multiple-input, multiple-output) control is devoted to precisely this topic.\cite{skogestad05}  There are several strategies that arise from this work, including optimal, robust, and adaptive control.\cite{skogestad05,astrom08}  Why not use those instead?  These strategies all require at least some systematic characterization of the transfer function of the system under control, which can be  time consuming.  They also require a sophisticated understanding of control theory, even if software implementations are now widely available.  Indeed, surveys report that nearly all industrial  applications continue to use PID feedback (e.g., 97\% of regulators in the refining, chemical, and pulp industries\cite{astrom08b}).  The split-PID strategy described here maintains the simplicity, intuitiveness, and ease of application of the usual PID strategy.  To modify a conventional controller, all that is needed is to add a second sensor and to alter a few lines of code in the controller software.  Finally, we emphasize that although we have considered temperature control as a specific case, PID controllers are used for many other applications, including control of flow, pressure, velocity, frequency, and more.  In all of these cases, the ideas presented here may be advantageous.
%In all of these cases, split-PID control may be advantageous.  [slightly shorter if we need to shorten that paragraph by one line].

\vspace{1em}
This work was funded by NSERC (Canada).  We thank Jeff Rudd for assembling the experimental  device and Normand Fortier for help with a preliminary version of the experiment.


\begin{thebibliography}{99}
% The numeral (here 99) in curly braces is nominally the number of entries in
% the bibliography. It's supposed to affect the amount of space around the
% numerical labels, so only the number of digits should matter--and even that
% seems to make no discernible difference.

\bibitem{bechhoefer05}  J. Bechhoefer, Rev. Mod. Phys. \textbf{77}, 783 (2005).

\bibitem{astrom08b} K. J. {\AA}str\"om and R. M. Murray, \textit{Feedback Systems:  An Introduction for Scientists and Engineers}, (Princeton Univ. Press, 2008).

\bibitem{skogestad05} S. Skogestad and I. Postlethwaite, \textit{Multivariable Feedback Control} (John Wiley and Sons, 2005).

%\bibitem{astrom06} K. J. {\AA}str\"om and T. Hagglund, \textit{Advanced PID Control} (ISA---The Instrumentation, Systems, and Automation Society, 2006).

%\bibitem{ho10} J. Ho, T. R. Jow, and S. Boggs, IEEE Electrical Insulation Magazine \textbf{26}, 20 (2010).

\bibitem{bechhoefer07} J. Bechhoefer, Y. Deng, J. Zylberberg, C. Lei, and Z.-G. Ye, Am. J. Phys. \textbf{75}, 1046 (2007).

\bibitem{astrom08} K. J. {\AA}str\"om and B. Wittenmark, \textit{Adaptive Control}, 2nd ed.  (Dover Publications, 2008).

\end{thebibliography}
\end{document}